\documentclass[aps,twocolumn]{revtex4} 

\usepackage[dvipdfmx]{graphicx} 
\usepackage{tabularx}
\usepackage{dcolumn} 
\usepackage{color}
\usepackage{bm}
\usepackage{amsmath}
\usepackage{amssymb}
\usepackage{here}
\usepackage{ulem}
\usepackage{physics}

\begin{document} 


\title{Discovery of antiferromagnetic chiral helical ordered state in trigonal GdNi$_3$Ga$_9$}


\author{Shota Nakamura$^{1}$, Takeshi Matsumura$^{2}$, Kazuma Ohashi$^1$, Hiroto Suzuki$^1$, Mitsuru Tsukagoshi$^{2}$, Kenshin Kurauchi$^{2}$, Hironori Nakao$^{3}$, Shigeo Ohara$^{1}$}
\affiliation{$^1$Department of  Physical Science and Engineering, Nagoya Institute of Technology, Nagoya 466-8555, Japan}
\affiliation{$^2$Department of Quantum Matter, ADSE, Hiroshima University, Higashihiroshima, Hiroshima 739-8530, Japan} 
\affiliation{$^3$Photon Factory, Institute of Materials Structure Science, High Energy Accelerator Research Organization, Tsukuba, 305-0801, Japan}

\date{\today}

\begin{abstract}

We have performed magnetic susceptibility, magnetization, and specific heat measurements on a chiral magnet GdNi$_3$Ga$_9$, belonging to the trigonal space group $R32$ (\#155). A magnetic phase transition takes place at $T_{\rm N}$~=~19.5~K. 
By applying a magnetic field along the $a$ axis at 2 K, the magnetization curve exhibits two jumps at $\sim 3$~kOe and =~45~kOe. 
To determine the magnetic structure, we performed a resonant X-ray diffraction experiment by utilizing a circularly polarized beam. 
It is shown that a long-period antiferromagnetic (AFM) helical order is realized at zero field. The Gd spins in the honeycomb layer are coupled in an antiferromagnetic manner in the $c$ plane and rotate with a propagation vector $\vb*{q}$ = (0, 0, 1.485). 
The period of the helix is 66.7 unit cells ($\sim 180$~nm). In magnetic fields above 3~kOe applied perpendicular to the helical $c$ axis, the AFM helical order changes to an AFM order with $\vb*{q}$ = (0, 0, 1.5).


\end{abstract}

\maketitle

\maketitle

\section{Introduction}
Asymmetry in space and time often plays an important role in unconventional physical phenomena.
Chirality is a property of asymmetry essential in cross-disciplinary subjects and nature at all large scales.
In structures with chirality,  there are two types of structures, i.e., left- and right-handed mirror copies.
The difference in chirality, or the handedness, of crystals often appears in macroscopic physical properties.
In magnets, chiral symmetry breaking in crystal structures plays an essential role in stabilizing a macroscopic ordering state with left- or right-handed incommensurate twists of magnetic spins.

Magnetic structures reflecting the crystal chiralities have been identified mainly in $d$ electron systems and have been actively studied.   The skyrmion lattice, which is a magnetic vortex, is observed in MnSi~\cite{muhlbauer2009skyrmion}, and a uniaxial helical magnetic structure with a periodic arrangement of "twists", which is called a chiral soliton lattice (CSL), is observed in CrNb$_3$S$_6$~\cite{togawa2012chiral,togawa2016symmetry}. 

\begin{figure}[!hbt]
\begin{center}
\includegraphics[width=1\linewidth]{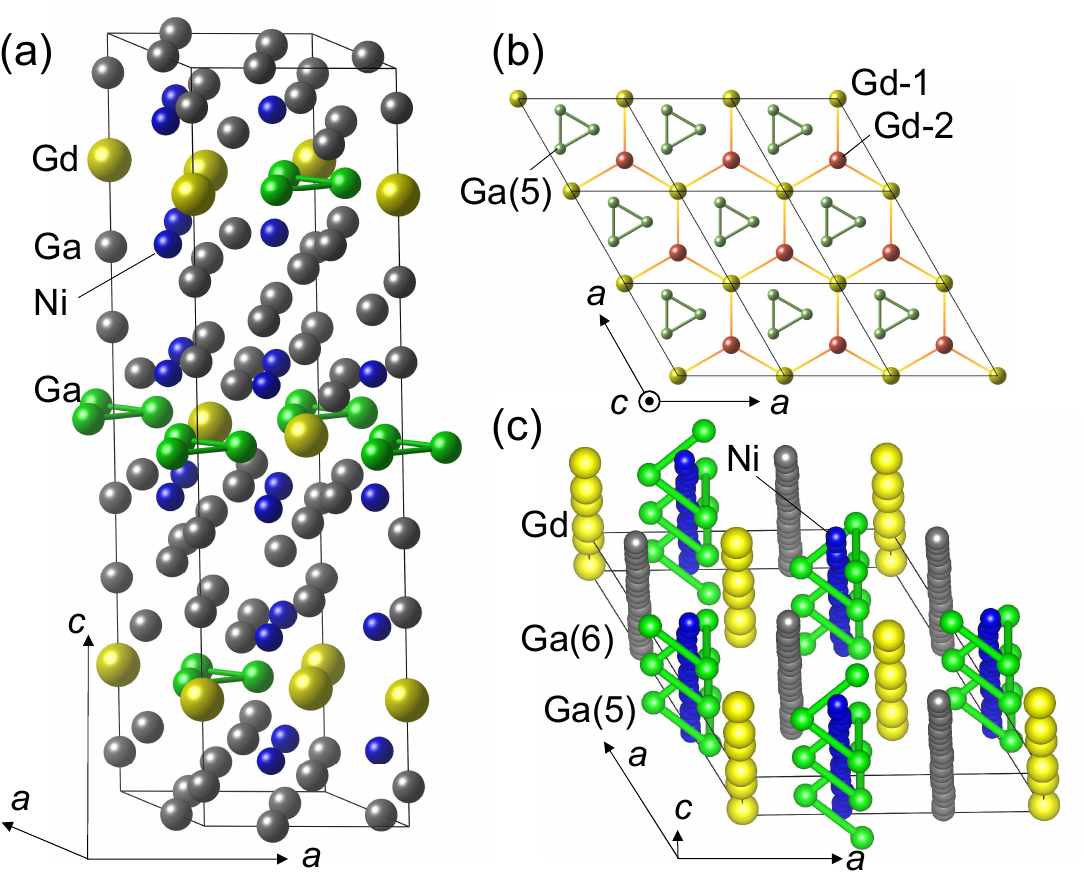}
\caption{(Color Online) 
(a) The unit cell of GdNi$_3$Ga$_9$. Ga(5), which is located in the Gd honeycomb layer, is colored green.  (b) Gd honeycomb layer ($z \sim 1/6$). This figure contains 9 unit cells. Gd forms a honeycomb network. Ga(5) forms a small triangle at the center of this network. (c)  Spiral in the crystal structure of GdNi$_3$Ga$_9$. The figure shows the $c$ axis spiral of Ga(5). Hard to see because the changes are so small, atoms in general positions, such as Ni and Ga(6), are also arranged in spirals in addition to Ga(5). In this figure, Ga(1)-Ga(4) was eliminated for simplicity. VESTA was used to draw the figure~\cite{momma2011vesta}.
}
\label{structure}
\end{center}
\end{figure}

Recently, chiral magnets have been explored in 4$f$ electron system.
The examples are uniaxial helical magnetic structures in YbNi$_3$Al$_9$~\cite{ohara2011transport,matsumura2017chiral}, DyNi$_3$Ga$_9$~\cite{ninomiya2017magnetic,ishii2018ferroquadrupolar,ishii2019anisotropic,tsukagoshi2022competition}, and GdPt$_2$B~\cite{sato2022new}, and the skyrmion lattices in EuPtSi~\cite{kakihana2018giant,kakihana2019unique,tabata2019magnetic,sakakibara2019fluctuation}.
The period of the helical magnetic structure in 4$f$ chiral magnet, for example in YbNi$_3$Al$_9$ and EuPtSi, is approximately an order of magnitude smaller than that in $d$ electron systems~\cite{togawa2016symmetry}.
 In the case of the helical magentic order of YbNi$_3$Al$_9$, the angle between neighboring spins along the $c$ axis are more than 90 degrees, indicating that the exchange interaction is mediated predominantly by the conduction electrons (the Ruderman-Kittel-Kasuya-Yosida mechanism), and the weak Dzyaloshinskii-Moriya (DM)-type antisymmetric exchange interaction selects the sense of rotation. 

\begin{figure}[t]
\begin{center}
\includegraphics[width=1\linewidth]{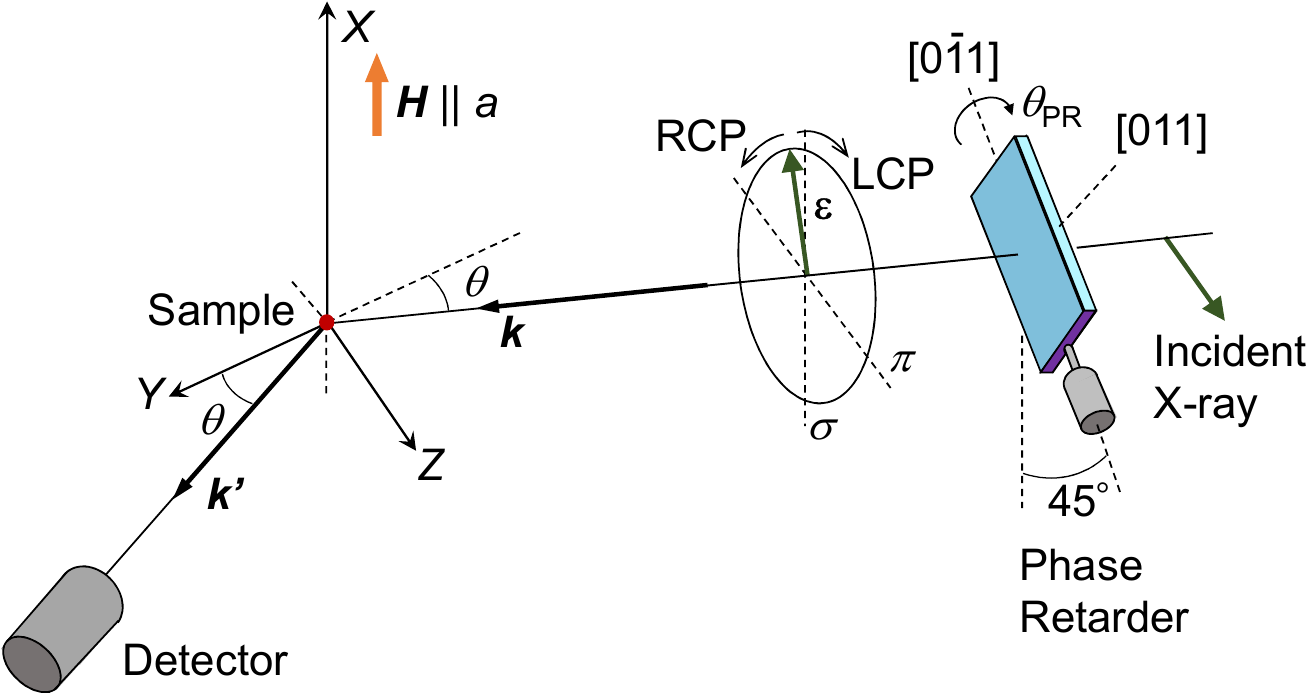}
\caption{(Color Online) 
Scattering configuration of the RXD experiments with a phase retarder system inserted in the incident x-ray. 
}
\label{Set}
\end{center}
\end{figure}
The target substance of this study is GdNi$_3$Ga$_9$~\cite{topertser2019crystal}, with the same crystal structure as YbNi$_3$Al$_9$, which has a ErNi$_3$Al$_9$-type crystal structure with a space group $R$32 (\#155)~\cite{nakamura2020enantiopure}. Figure \ref{structure} shows the crystal structure of GdNi$_3$Ga$_9$. We call this structure as right-handed and the mirror image of this structure as left-handed.  
The handedness can be judged by the Flack parameter of single crystal x-ray experiments.
The lattice constants are $a$~$=$~0.7264 and $c$~$=$~2.7497~nm.
The crystal structure of GdNi$_3$Ga$_9$ is characterized by the honeycomb layers of rare earth elements stacked along the helical $c$ axis, as shown in Figs. \ref{structure}(a) and \ref{structure}(b).
Note that the Sohncke type space group $R$32 is not chiral, but the crystal structure has a chirality.
Ga and Ni atoms form spirals, as shown in Fig. \ref{structure}(c).

Gd compounds are suitable for investigating chiral magnetism because of the zero orbital angular momentum of  Gd$^{3+}$ ($4f^7$, $^8{\rm S}_{7/2}$, $\mu_{\rm eff} = 7.94\mu_{\rm B}/{\rm Gd}$, $\mu_J = g_J\mu_{\rm B}J = 7\mu_{\rm B}$, and $g_J = 2$). 
This reduces the effects of crystal-field anisotropy and allows the magnetic moments of Gd to direct to any favorable directions determined by the exchange interactions.

Although neutron diffraction experiments are usually used to investigate the magnetic structure~\cite{ishida1985crystal,yamasaki2007electric}, detailed data collection for Gd compounds is difficult because Gd is a strong neutron absorber.
To observe the helical magnetic structure and the helicity of GdNi$_3$Ga$_9$, we performed resonant x-ray diffraction (RXD) using circularly polarized x-rays \cite{sutter1997helicity,PhysRevLett.102.237205,sagayama2010observation}.

\section{Experimental Procedure}
We synthesized a single crystalline GdNi$_3$Ga$_9$ by the flux method using gallium as the solvent.
The starting materials Gd ingot (purity of 99.9~\%), Ni ingot (99.999~\%), and Ga ingot (99.9999~\%) were prepared in a molar ratio of Gd: Ni: Ga = 1: 3: 30. 
These materials were placed in an alumina crucible and then sealed into a quartz tube under high vacuum $\sim 1 \times 10^{-3}$~Pa.
The ampule was heated to 900$^\circ$C and maintained at this temperature for 5 hours.
Then, it was slowly cooled at a rate of 5$^\circ$C/h.
The excess gallium was removed by centrifugation at 500~$^\circ$C.

In this work, the magnetization and specific heat measurements have been performed using a commercial magnetic properties measurement system (MPMS, Quantum Design) and physical properties measurement system (PPMS, Quantum Design). The specific heat was measured between 2 and 300~K at zero magnetic field, and the magnetization was measured between 2 and 300~K in magnetic fields up to 5.5~T. 

The RXD experiments were performed at BL-3A, Photon Factory at KEK, Japan.
The sample with a mirror polished $c$ plane surface was inserted into an 8~T cryo-magnet and cooled down to 3~K.
In this paper, the $c$ plane denotes the (001) plane.
The $c$ axis was perpendicular to the magnetic field.
The scattering geometry is shown in Fig. \ref{Set}. A diamond phase retarder system was used to tune the horizontally polarized incident beam to a circularly polarized state~\cite{matsumura2017chiral}. 
A phase difference of the incident x-ray between the $\sigma$ and $\pi$ polarization is generated by rotating the angle of the diamond phase plate $\theta_{\rm PR}$ around the 111 Bragg angle $\theta_{\rm B}$, which is 22.3$^\circ$ at 7.932~keV, where the scattering plane is tilted by 45$^\circ$.
We can tune the incident linear polarization to right-handed circular polarization (RCP) and left-handed circular polarization (LCP) by changing $\Delta\theta_{\rm PR}$, where $\Delta\theta_{\rm PR}~=~\theta_{\rm PR}-\theta_{\rm B}$.


We used an x-ray beam with a wavelength of 0.1711~ and 0.1563~nm  (7.246 and 7.932 keV), which are near the $L_3$ and $L_2$ edges of Gd, respectively. 
In GdNi$_3$Ga$_9$, the (1, 0, 3$n$+1) reflections are allowed and the (1, 0, 3$n$) and (1, 0, 3$n$$-$1) reflections are forbidden, because of the selection rule derived from the crystal symmetry.

\section{Results and Analysis}

\begin{figure*}[!hbt]
\begin{center}
\includegraphics[width=1\linewidth]{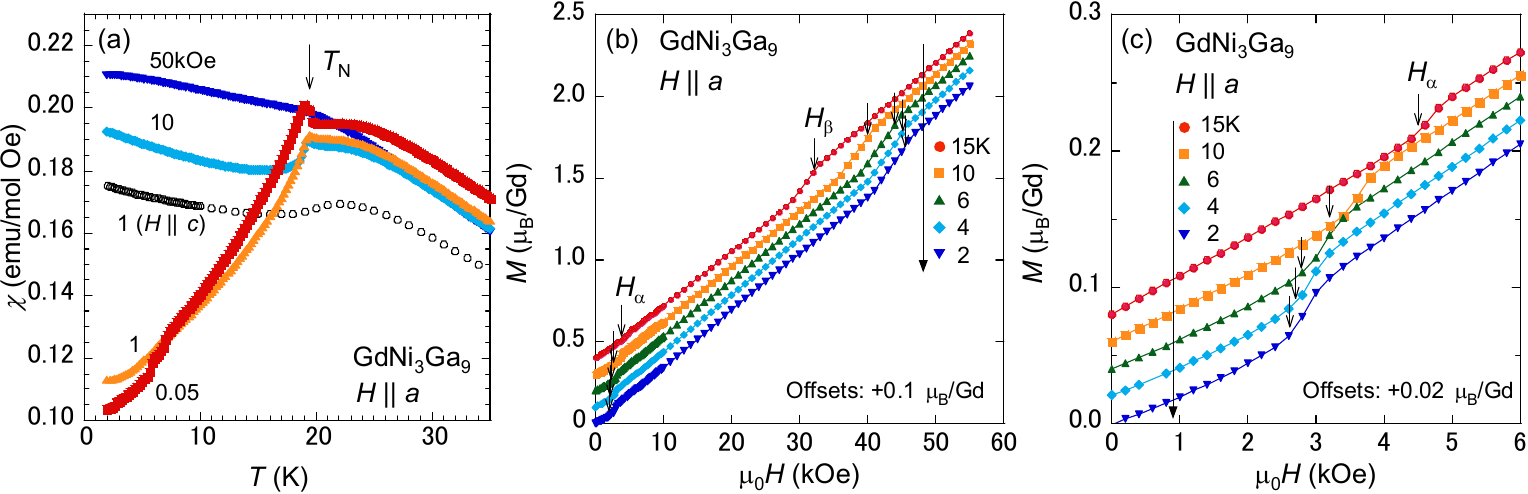}
\caption{(Color Online) 
(a) Temperature dependences of magnetic susceptibility obtained in the magnetic fields of 0.05, 1, 10, and 50 kOe along the $a$ axis, together with that obtained in  $H \parallel c$ at 1 kOe. (b) Magnetization curves were measured at 2, 4, 6, 10, and 15 K with increasing the magnetic field. (c) Enlarged view of low magnetic field region of Fig. \ref{chiT}(b).
}
\label{chiT}
\end{center}
\end{figure*}
\begin{figure}[t]
\begin{center}
\includegraphics[width=0.75\linewidth]{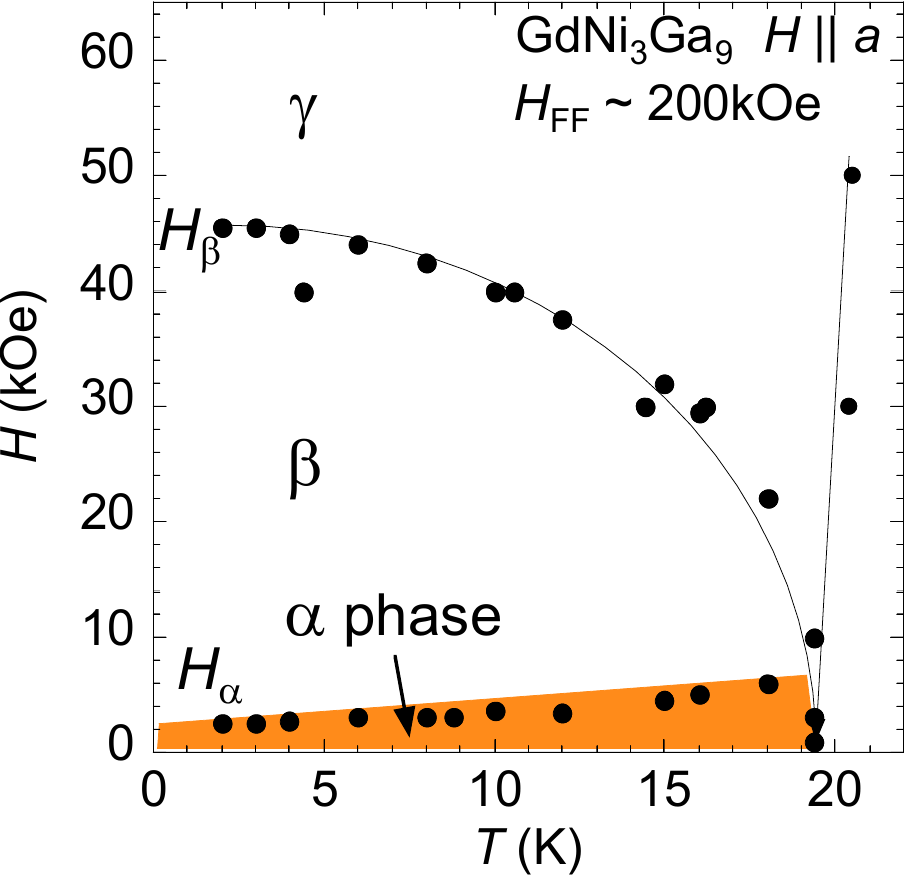}
\caption{(Color Online) 
$H-T$ phase diagram of GdNi$_3$Ga$_9$ in $H~\parallel~a$. The dots in this figure are obtained by magnetization measurements. The $\alpha$ phase is the AFM helical phase.
}
\label{PD}
\end{center}
\end{figure}
\begin{figure}[t]
\begin{center}
\includegraphics[width=0.8\linewidth]{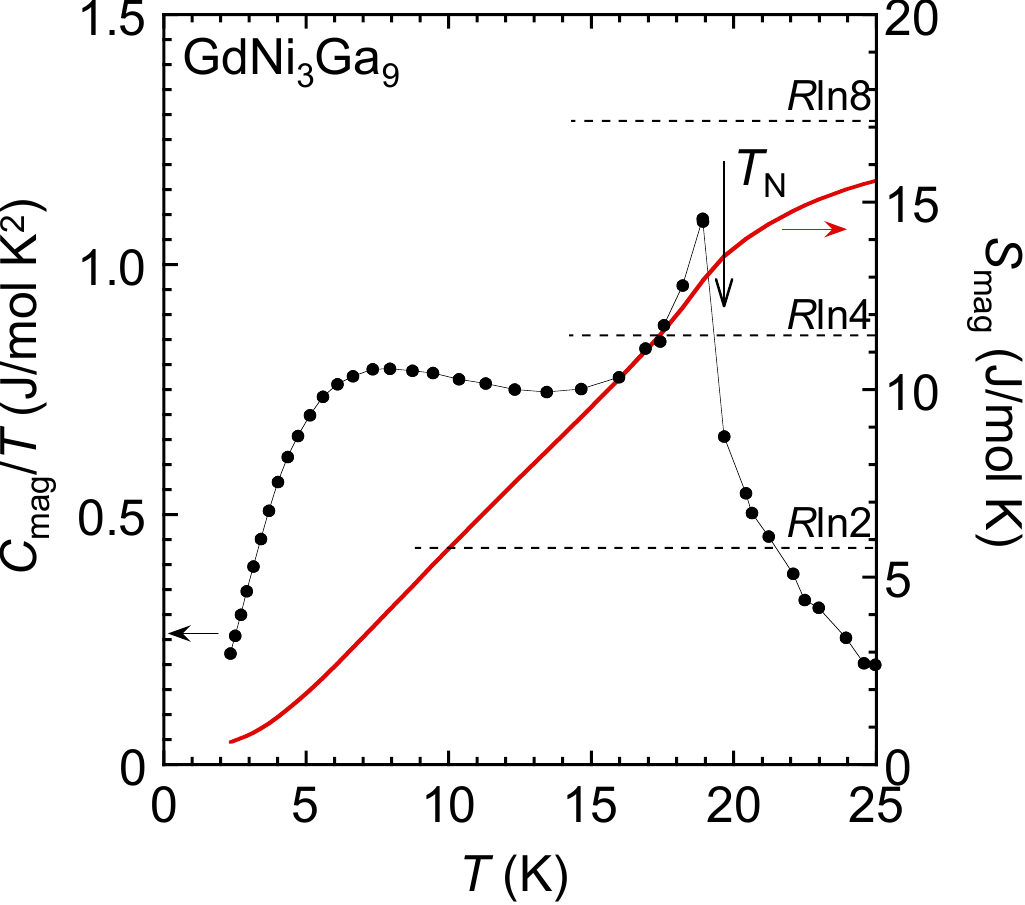}
\caption{(Color Online) 
Temperature dependence of specific heat divided by temperature $C_{\rm mag}/T$, together with the entropy $S_{\rm mag}$.
}
\label{C-T}
\end{center}
\end{figure}

Figure \ref{chiT}(a) shows the temperature dependence of magnetic susceptibility ($\chi$$-$$T$) of GdNi$_3$Ga$_9$, which was measured at 0.05, 1, 10, and 50 kOe applied along the $a$ axis ($H \parallel a$).
In this figure, $\chi$$-$$T$ for $H \parallel c$ at 1 kOe is also shown. 
This result suggests that the Gd spins in a honeycomb layer of Fig. \ref{structure}(b) are antiparallel to each other below $T_{\rm N} \sim$~19.5~K.
 As shown afterwards, this material exhibits antiferromagnetic (AFM) helical order where magnetic propagation vector $\vb*{q}~\parallel~c^*$.
At low magnetic fields of 0.05 and 1~kOe, a characteristic cusp is observed just below $T_{\rm N}$, and $\chi$ decreases monotonically with decreasing temperature.
This cusp is similar to the one theoretically predicted to appear when the chiral helix is realized~\cite{kishine2005synthesis,kousaka2007chiral}.
This anomaly is not observed in $H \parallel c$.
As the magnetic field is increased, the behavior of $\chi$$-$$T$ curve changes below $T_{\rm N}$.
At 10~kOe, $\chi$ decreases below $T_{\rm N}$ and then increases slightly at low temperatures.
At 50~kOe, bending of $\chi$ due to AFM ordering becomes gradual.
Warming from $T_{\rm N}$ to room temperature, $\chi$$-$$T$ curve of this material  is in good agreement with the Curie-Weiss law, and the Curie-Weiss temperature is estimated to be $-$8.7 and $-$11.2~K for the $a$ and $c$ axes, respectively. 
Although not shown in this paper, this material is isotropic in the $c$ plane, and similar magnetization curves are obtained when a magnetic field is applied in the $a^*$ direction.

Figure \ref{chiT}(b) shows magnetization curves measured at 2, 4, 6, 10, and 15~K.
All magnetization curves jump twice at $H_{\alpha} \sim 3$~kOe and $H_{\beta} \sim 45$~kOe.
Except for these two jumps, magnetization increases linearly.
With increasing temperature, $H_{\alpha}$ and $H_{\beta}$ shift toward higher and lower magnetic fields, respectively.
Figure \ref{chiT}(c) shows an enlarged view of the low magnetic field region in Fig. \ref{chiT}(b).
In this figure, we can see the shift of $H_{\alpha}$ to the high field side with increasing temperature.
The metamagnetic behavior at critical magnetic fields during the linearly increasing magnetization process is a feature that appears in the magnetization process of the AFM helix when a magnetic field is applied perpendicular to the helix axis.~\cite{kishine2005synthesis,kousaka2007chiral}

Figure \ref{PD} shows the $H-T$ phase diagram of GdNi$_3$Ga$_9$ for $H \parallel a$ obtained by the magnetization measurements.
The magnetically ordered phases are named $\alpha$, $\beta$, and $\gamma$ phases in the order from low to high fields.
$H_{\alpha}$ and $H_{\beta}$ are located between the $\alpha$ and $\beta$ phases, and the $\beta$ and $\gamma$ phases, respectively.
$\gamma$ phase changes to the forced-ferromagnetic state at $H_{\rm FF} \sim$ 200 kOe, when we assume that the magnetization curve of Fig. \ref{chiT}(b)   behaves linearly up to 7$\mu_{\rm B}$, the saturation magnetization of Gd$^{3+}$.
As will be discussed later, the $\alpha$ phase is an AFM helical phase, and the $\beta$ and $\gamma$ phases are an AFM phase.

Figure \ref{C-T} shows the temperature dependence of specific heat divided by temperature ($C_{\rm mag}/T$) at zero field, together with the entropy ($S_{\rm mag}$).
$C_{\rm mag}/T$ is obtained by subtracting the specific heat of nonmagnetic LuNi$_3$Ga$_9$.
 $S_{\rm mag}$ is obtained from $C_{\rm mag}/T$ assuming that $C_{\rm mag}/T$ increases linearly from ($T$, $C_{\rm mag}/T$)~=~(0, 0).
A peak of $C_{\rm mag}/T$ appears just below $T_{\rm N} \sim 19.5$~K.
The entropy is estimated to be $\sim$~$R{\rm ln4}$ at $T_{\rm N}$.
Below $T_{\rm N}$, a shoulder peak is observed around 7~K.
Such a shoulder has often been observed at around $T_{\rm N}$/4 in Gd$^{3+}$ compounds with $J$ = 7/2~\cite{hidaka2020helical,PhysRevB.43.13137}, and it is attributed to a  splitting of the $J$ = 7/2 multiplet due to an internal field~\cite{PhysRevB.43.13145}. 

\begin{figure}[t]
\begin{center}
\includegraphics[width=1\linewidth]{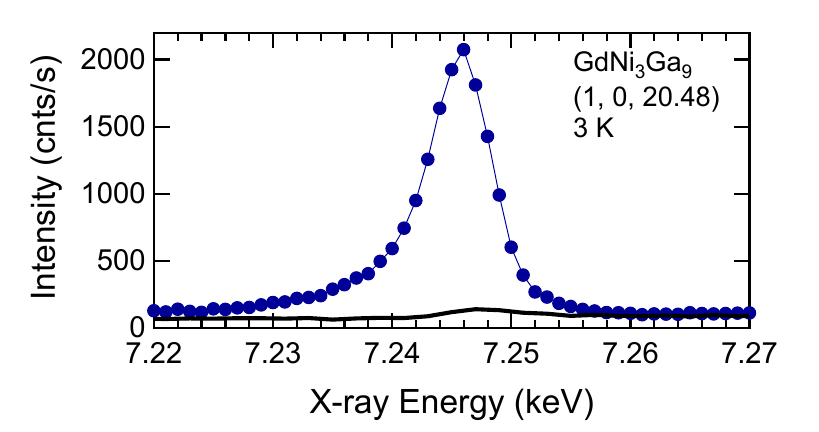}
\caption{(Color Online) 
X-ray energy $E$  dependence of the (1, 0, 20.48) reflection around $E = 7.246$~keV, together with the background (black curve) obtained with $\theta$ shifted slightly from (1, 0, 20.48).
}
\label{Escan}
\end{center}
\end{figure}

\begin{figure}[t]
\begin{center}
\includegraphics[width=1\linewidth]{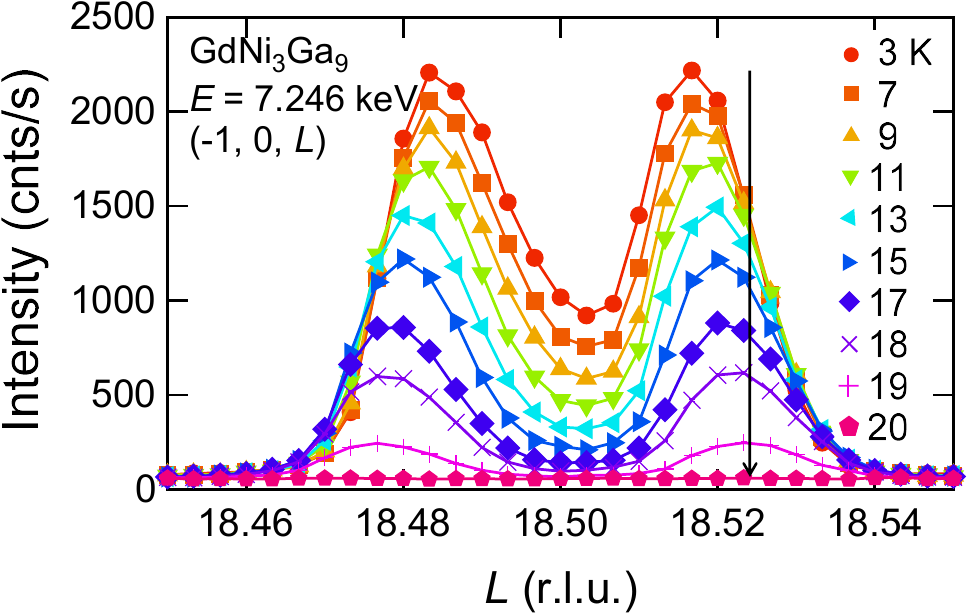}
\caption{(Color Online) 
Temperature dependence of the ($-$1, 0, $L$) peak profile at zero magnetic field in the AFM helical $\alpha$ phase.
}
\label{Tscan}
\end{center}
\end{figure}
\begin{figure}[t]
\begin{center}
\includegraphics[width=0.85\linewidth]{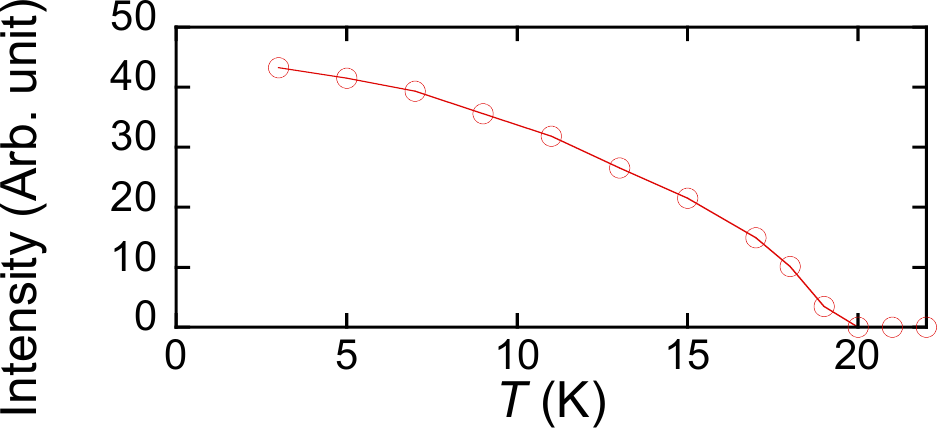}
\caption{(Color Online) Temperature variation of the ($-$1, 0, 17+$q$) peak intencity shown in Fig. \ref{Tscan}.
}
\label{Tscanpic}
\end{center}
\end{figure}

To observe the helical magnetic structure and the magnetic helicity of GdNi$_3$Ga$_9$, a RXD study has been performed.
The measurement was conducted for the right-handed crystal, whose structure is shown in Fig. \ref{structure}.
The crystal chirality is reflected in helicity of the magnetic structure by a one-by-one relationship in YbNi$_3$Al$_9$~\cite{matsumura2017chiral}, and left-handed crystals are expected to describe the inverted helix of the magnetic structure of this paper.

We found the resonant Bragg reflection from magnetic order at (1, 0, $3n+1\pm q$) and ($-$1, 0, $3n-1\pm q$) with $q~=~1.485$ at 3~K. 
The X-ray energy dependence for the (1, 0, 20.48) peak is shown in  Fig. \ref{Escan}. 
The intensity enhancement due to the $E1$ resonance ($2p~\leftrightarrow~5d$) is observed around the $L_3$-edge of Gd, indicating that the resonant signal reflects the magnetic ordering of the Gd $4f$-spins.  
The peak position of the intensity is 7.246 keV. 
It is noted that no reflection was found along the (0, 0, $L$) line at (0, 0, $3n \pm q$), indicating that the magnetic moments of the two Gd spins on the same honeycomb layer are antiparallel.
The detail will be explained later.

Figure \ref{Tscan} shows the temperature dependence of the ($-$1, 0, $L$) peak profile at zero magnetic field in the AFM helical phase.
The two peaks at ($-$1, 0, 18.485) and ($-$1, 0, 18.515) at 3 K correspond to ($-$1, 0, 17+$q$) and ($-$1, 0, 20$-q$) with $q = 1.485$. 
The peak intensity decreases with increasing $T$, and disappears at $T_{\rm N}$ as shown in Fig. \ref{Tscanpic}. 
As explained later, the magnetic moments lie in the $c$ plane and exhibit a helical rotation along the $c$ axis. 
The $q$-value of 1.485 means that the spins rotate by 178.2$^\circ$ between the adjacent Gd honeycomb layers, and 360$^\circ$ with 0.673 unit cells (approximately two Gd honeycomb layers).
The deviation of the angle from 180$^\circ$ gives rise to the twist angle of the helix of 1.8$^\circ$ corresponding to $q$ = 0.015, which is the difference between 1.485 and 1.5.
The period of the AFM helix with the twist angle of 1.8$^\circ$ is 66.7 unit cells ($\sim$~180 nm, 200 layers), which is required to come back to the same direction as the original layer.

This long period of the helix is comparable with those of chiral magnets of $d$ electron systems. 
The magnetic propagation vector decreases slightly to $\vb*{q}~=$~(0, 0, 1.48) by increasing temperature to $T_{\rm N}$. When $\vb*{q}~=$~(0, 0, 1.48), the period of the AFM helix becomes 3/4 times shorter than the one at 3 K, indicating that the AFM helix is more twisted at higher temperatures.

\begin{figure}[t]
\begin{center}
\includegraphics[width=1\linewidth]{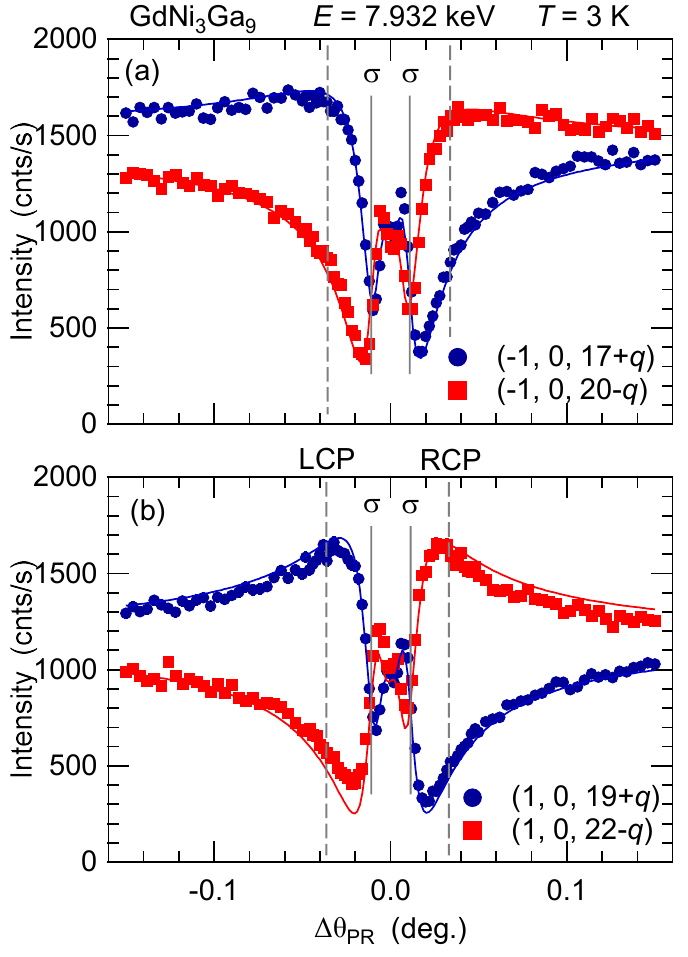}
\caption{(Color Online) 
$\Delta\theta_{\rm PR}$ scans, which show the intensity relations for the RCP and LCP incident x-rays at (a) ($-$1, 0, 17+$q$) and ($-$1, 0, 20$-q$), (b) (1, 0, 19+$q$) and (1, 0, 22$-q$) reflection peaks.
}
\label{Xalpha}
\end{center}
\end{figure}

To show that the helical order has a unique sense of rotation, which should be determined by the antisymmetric interaction, we investigated the intensity variation by manipulating the circular polarization of the incident beam. 
The RCP and LCP are obtained by changing $\Delta\theta_{\rm PR}$ = $\theta_{\rm PR} - \theta_{\rm B}$, where $\theta_{\rm PR}$ and $\theta_{\rm B}$ denote the angle of the diamond phase plate and the 111 Bragg angle, respectively. It is convenient to express the polarization state of the X-ray photon by the Stokes vector $\vb*{P}$ = ($P_1$, $P_2$, $P_3$), where $P_1$, $P_2$, and $P_3$ denote the degrees of $\pm$45$^{\circ}$ ($P_1$ = $\pm$1) linear polarization, the RCP ($P_2$ = 1) or LCP ($P_2$ = $-$1), and $\sigma$ ($P_3$ = 1) or $\pi$ ($P_3$ = $-$1) linear polarization state, respectively. 
In our geometry shown in Fig. \ref{Set}, the three parameters are expressed as $P_1$~=~0, $P_2$~=~sin(A/$\Delta\theta_{\rm PR}$) and $P_3$~=~$-$cos(A/$\Delta\theta_{\rm PR}$), where A is a constant determined experimentally. 
The LCP and RCP are obtained when $\Delta\theta_{\rm PR}$ = -0.0358 and 0.0358 deg., respectively, whose positions are indicated by vertical dotted lines in Fig. \ref{Xalpha}.

\begin{figure}[t]
\begin{center}
\includegraphics[width=0.95\linewidth]{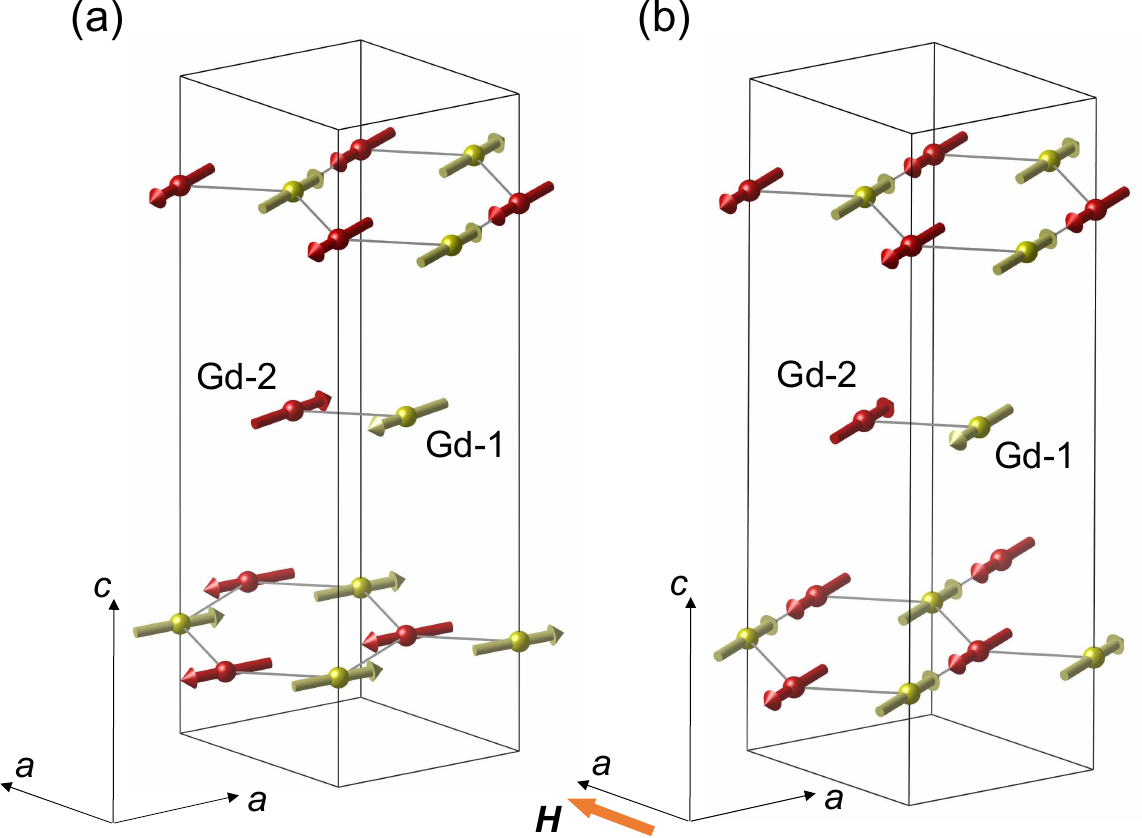}
\caption{(Color Online) 
Schematic view of the AFM helical and commensurate AFM structures in the $\alpha$ and $\beta$ phases of GdNi$_3$Ga$_9$, respectively. In this figure, the twist angle is emphasized about 10 times. VESTA was used to draw the figure~\cite{momma2011vesta}.
}
\label{MS}
\end{center}
\end{figure}

Figure \ref{Xalpha} shows the $\Delta\theta_{\rm PR}$ scans, which demonstrate that this material has the helical magnetic structure in the $\alpha$ phase.
 The $\Delta\theta_{\rm PR}$ dependences for the  ($-$1, 0, 17+$q$)  and the  ($-$1, 0, 20$-q$) peaks are opposite to each other, indicating that the magnetic structure in the $\alpha$ phase has a chirality. The intensity of the ($-$1, 0, 17+$q$) peak is larger (smaller) than that of ($-$1, 0, 20$-q$) for LCP (RCP). 
The same result is obtained for (1, 0, 19+$q$) and (1, 0, 22$-q$) as well. The solid lines in Fig. \ref{Xalpha}  are the calculated curves of the $\Delta\theta_{\rm PR}$ scans by assuming an AFM helical magnetic structure as described below. This calculation agrees well with the $\Delta\theta_{\rm PR}$ scan data.

Let us analyze the magnetic structure in the $\alpha$ phase. The magnetic moments of Gd, which are antiferromagnetically coupled in the honeycomb layer in the $c$ plane, are helically twisted along the $c$ axis with a very long pitch.  
There are two Gd atoms for the $6c$ site of the unit cell of GdNi$_3$Ga$_9$ with a space group of $R32$: Gd-1 at $\vb*{d}_1$ = (0, 0, $z$), and Gd-2 at $\vb*{d}_2$ = (0, 0, $\bar{z}$), where $z~\sim$~1/6.
In the present case, the magnetic moments are expected to be antiparallel to each other in the $c$ plane. 
There are six Gd positions in the unit cell, as indicated by the Wyckoff letter $6c$ of Gd in GdNi$_3$Ga$_9$. The set of Gd-1 and Gd-2 are chosen to be closest to each other, and the one-to-three assignments are determined, where positions are connected by the stacking vector (2/3, 1/3, 1/3) each other.
The unit vector $\vb*{\mu}_{1, j}$ and $\vb*{\mu}_{2, j}$ parallel to magnetic moments of Gd-1 and Gd-2, respectively, on the $j$-th lattice point at $\vb*{r}_j$~=~($n_1$, $n_2$, $n_3$), ($n_1$+2/3, $n_2$+1/3, $n_3$+1/3), and ($n_1$+1/3, $n_2$+2/3, $n_3$+2/3), where $n_1$, $n_2$, and $n_3$ are integers, are expressed as 
\begin{flalign}
&\vb*{\mu}_{1, j} = \hat{\vb*{x}}~{\rm cos}2\pi\vb*{q} \cdot \vb*{r}_j+\hat{\vb*{y}}~{\rm cos}(2\pi\vb*{q} \cdot \vb*{r}_j+\varphi),\\
&\vb*{\mu}_{2, j} = \hat{\vb*{x}}~{\rm cos}(2\pi\vb*{q} \cdot \vb*{r}_j+\delta)+\hat{\vb*{y}}~{\rm cos}(2\pi\vb*{q} \cdot \vb*{r}_j+\varphi+\delta).
\end{flalign}
The $\hat{\vb*{x}}$ and $\hat{\vb*{y}}$ denote the unit vectors along the  $x$- and $y$-axis, respectively, which are perpendicular to the  $c$ ($z$) axis. 
$\varphi$ takes values of $\pm\pi$/2, and describes a right- or left-handed helical structure. 
When $\delta$~=~178.2$^\circ$, the two spins on the same layer are perfectly antiferromagnetic for $q$~=~1.485. 
Note that $\vb*{\mu}_{1, j}$ and $\vb*{\mu}_{2, j}$ are different layers, and they have a relation of $\vb*{\mu}_{1, j} = - \vb*{\mu}_{2, j+1}$ on the same layer.
A schematic view of the magnetic structure in the $\alpha$ phase is shown in Fig. \ref{MS}(a).  The twist angle of the helix is emphasized by 10 times.

The calculated $\Delta\theta_{\rm PR}$ dependence of the intensity for the above AFM-helical model well explains the data in Fig. \ref{Xalpha}, which were obtained at the Gd $L_2$ edge.
The $E1$ resonant scattering amplitude from the magnetic dipole order is proportional to $i(\vb*{\varepsilon}' \times \vb*{\varepsilon}) \cdot \vb*{Z}_{\rm dip}^{(1)}$~\cite{PhysRevLett.61.1245,lovesey2005electronic}, where
\begin{align}
\vb*{Z}_{\rm dip}^{(1)}~=~\sum_{d, j} \vb*{\mu}_{d, j}e^{-i2\pi\vb*{Q} \cdot (\vb*{r}_j +\vb*{d})}
\end{align} 
represents the magnetic dipolar structure factor at the scattering vector $\vb*{Q} = \vb*{k}' - \vb*{k}$. 
Here, $\vb*{\varepsilon}$ and $\vb*{\varepsilon}'$ denote the polarization vector of the incident and diffracted beam, respectively. 
In the horizontal scattering plane configuration in this experiment, the incident linear polarization is $\pi$ when $\abs{\Delta\theta_{\rm PR}}$ is large.
At ($-$1, 0, 17+$q$), $\vb*{Z}_{\rm dip}^{(1)}$ is calculated to be (1, $\mp i$, 0) for $\varphi = \pm\pi$/2.
The sign is reversed at ($-$1, 0, 20$-q$), where $\vb*{Z}_{\rm dip}^{(1)}$~=~(1, $\pm i$, 0) for $\varphi = \pm\pi$/2.
For (0, 0, 3n+-q), $\vb*{Z}_{\rm dip}^{(1)}$ vanishes when the two Gd moments on the same layer are antiferromagnetic by setting $\delta$~=~178.2$^\circ$.


We use the scattering-amplitude-operator method to analyze the polarization properties of the experimental results~\cite{Lovesey}.
We consider a $2 \cross 2$ matrix $\hat{F}$, consisting of four elements of the scattering amplitudes for the polarization arrangements $\sigma-\sigma'$, $\pi-\sigma'$, $\sigma-\pi'$, and $\pi-\pi'$: 
\begin{align}
\hat{F}~=~\mqty(F_{\sigma\sigma'} & F_{\pi\sigma'} \\ F_{\sigma\pi'} & F_{\pi\pi'}).
\end{align} 
As shown in Appendix B in Ref. \cite{matsumura2017chiral}, the scattering cross section (d$\sigma$/d$\Omega$) is expressed as
\begin{equation}
\begin{split}
(\frac{d\sigma}{d\Omega}) &= \frac{1}{2}(|F_{\sigma\sigma'}|^2+|F_{\sigma\pi'}|^2+|F_{\pi\sigma'}|^2+|F_{\pi\pi'}|^2)\\
&+P_1~{\rm Re}\{F_{\pi\sigma'}^*F_{\sigma\sigma'}+F_{\pi\pi'}^*F_{\sigma\pi'}\}\\
&+P_2~{\rm Im}\{F_{\pi\sigma'}^*F_{\sigma\sigma'}+F_{\pi\pi'}^*F_{\sigma\pi'}\}\\
&+\frac{1}{2}P_3(|F_{\sigma\sigma'}|^2+|F_{\sigma\pi'}|^2-|F_{\pi\sigma'}|^2-|F_{\pi\pi'}|^2)\\
&= C_0 + C_1P_1 + C_2P_2 + C_3P_3, 
\end{split}
\end{equation}
where $P_1$~=~0 in the present geometry of the phase retarder. 
By using the scattering amplitude matrix $\hat{F}$~=~$i(\vb*{\varepsilon}' \times \vb*{\varepsilon}) \cdot \vb*{Z}_{\rm dip}^{(1)}$,  ($C_2$/$C_0$, $C_3$/$C_0$) is calcurated to be ($-$0.574, $-$0.562) and (0.574, $-$0.562) for ($-$1, 0, 17 + $q$) and ($-$1, 0, 20 - $q$), respectively, where $\varphi = \pi/2$. The solid lines in Fig. \ref{Xalpha} are the calculated curves using the above-mentioned magnetic structure model and agree with the experimental data.




\begin{figure}[t]
\begin{center}
\includegraphics[width=1\linewidth]{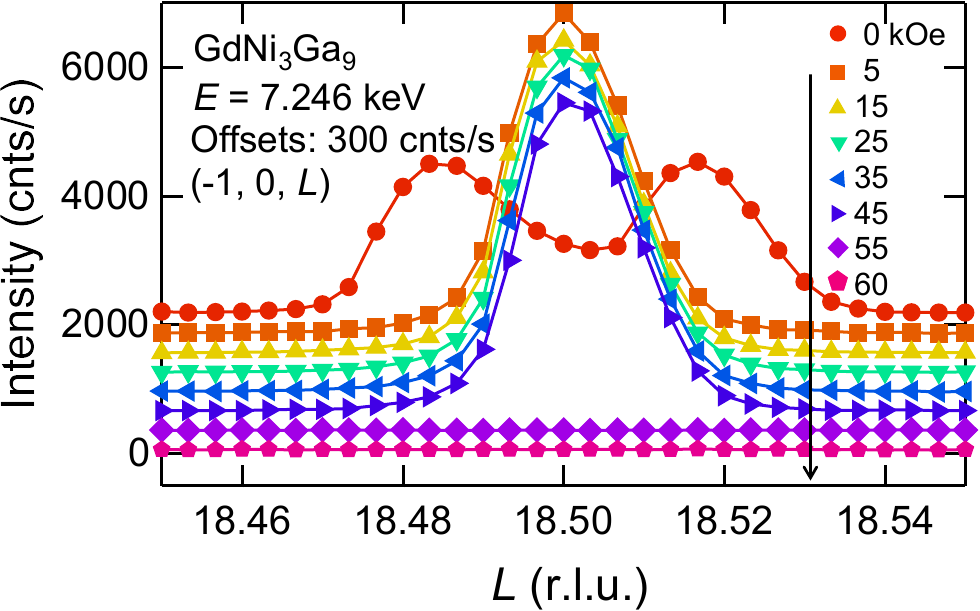}
\caption{(Color Online) 
Magnetic field dependence of the ($-$1, 0, $L$) peak profile at the lowest temperature 3~K.
}
\label{Hscan}
\end{center}
\end{figure}
\begin{figure}[t]
\begin{center}
\includegraphics[width=1\linewidth]{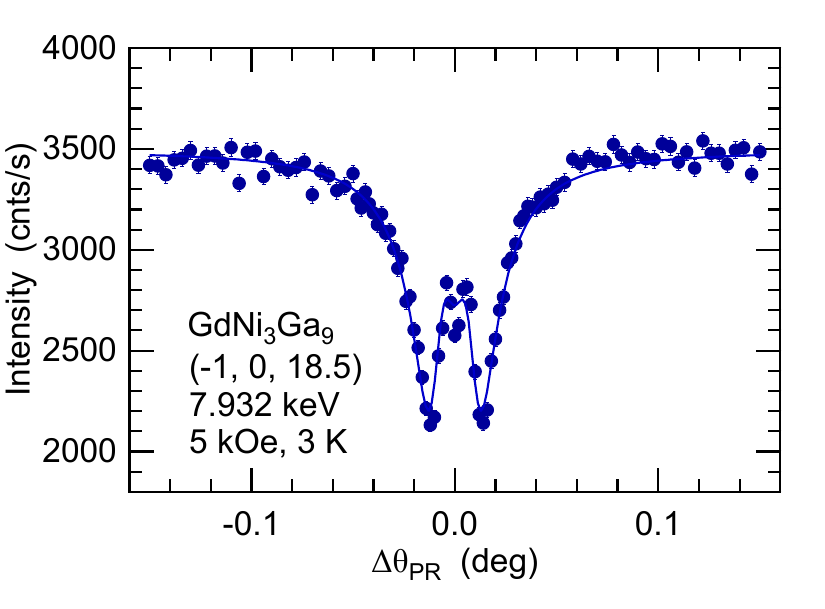}
\caption{(Color Online) 
$\Delta\theta_{\rm PR}$ scan at ($-$1, 0., 18.5), where $\vb*{q}~=$~(0, 0, 1.5), in the $\beta$ phase.
}
\label{Xbeta}
\end{center}
\end{figure}

The AFM-helical order soon changes to an AFM order by applying a magnetic field in the $c$ plane, where the AFM moments are aligned perpendicular to the field. 
Figure \ref{Hscan} shows the magnetic field dependence of the ($-$1, 0, $L$) peak profile at the lowest temperature of 3~K.
With increasing magnetic field, the AFM chiral helix peak soon moves to the commensurate position at $\vb*{q}~=$~(0, 0, 1.5) in a weak magnetic field of $\sim$~3~kOe above which the $\beta$ phase is realized.
The propagation vector $\vb*{q}~=$~(0, 0, 1.5)  corresponds to the antiferromagnetic alignment of adjacent Gd honeycomb layers.
The $q$-value sharply changes from 1.485 to 1.5 with increasing field, and it is hard to observe the soliton lattice~\cite{zheludev1997field}.
When a magnetic field is further increased, the transition from the $\beta$ to $\gamma$ phases occurs at 45~kOe. 
Although the magnetic Bragg peak was searched along the ($-$1, 0, $L$) line and several other conditions in the $\gamma$ phase, the peak could not be found. 
This result suggests that the $q$-vector may have changed from (0, 0, 1.5) to (0, 0, 0). 
The details of the $\gamma$ phase are still unknown. From the results of the magnetization measurements, however, the $M(H)$ curve suggests an AFM component remains.
ErNi$_3$Ga$_9$ and DyNi$_3$Ga$_9$ having the same crystal structure as this material also exhibit AFM order~\cite{tsukagoshi2022competition, ninomiya2018neutron}.

Figure \ref{Xbeta} shows the $\Delta\theta_{\rm PR}$ scan at ($-$1, 0, 18.5), where $\vb*{q}~=$~(0, 0, 1.5), in the $\beta$ phase.
The shape is different from the one in the $\alpha$ phase with $q$ = 1.485, and symmetrical around $\Delta\theta_{\rm PR}$ = 0, because of the loss of the chirality.
The solid line is a calculated one as in Fig. \ref{Xalpha} for a magnetic structure  as shown in Fig. \ref{MS}(b), where the antiferromagnetically-aligned magnetic moments of Gd orient perpendicular to the magnetic field direction $H \parallel a$ in the Gd honeycomb plane.
The experimental data are in good agreement with the calculation.

\section{Conclusion}
We have performed the magnetization and specific heat measurements, and resonant x-ray diffraction experiments to investigate the helical magnetic structure of GdNi$_3$Ga$_9$ with the Sohncke space group $R32$, where a chiral magnetic structure is allowed because of lacks both of inversion and mirror symmetries.
This material exhibits an AFM helical order at 19.5K, where a cusp in the magnetic susceptibility, which is a character of a chiral helix, is observed.
In this state, the magnetic moments of Gd twist  178.2$^\circ$ per a Gd honeycomb layer along the helical $c$ axis with $\vb*{q}~=$ (0, 0, 1.485), where the period of the helix is the length of 66.7 unit cells, and antiferromagnetically align in the $c$ plane.
We confirmed that the magnetic Bragg peaks with $\vb*{q}~=$ (0, 0, 1.485) exhibit opposite response to the RCP and LCP incident x-rays, indicating that the AFM helix consists of a single sense of rotation. 
With increasing magnetic field, the AFM helical state changes to an antiferromagnetic phase with $\vb*{q}~=$~(0, 0, 1.5) in $H \parallel a$ of $\sim$ 3~kOe. 
This $\vb*{q}$ corresponds to a magnetic structure where the magnetic moments between the adjacent Gd honeycomb layers are antiferromagnetically
aligned perpendicular to the $c$ axis.

GdNi$_3$Ga$_9$ and YbNi$_3$Al$_9$ have the same crystal structure and antiferromagnetic and ferromagnetic spiral magnetic structures, respectively.
In the future, research that compares these materials is desired.

\section*{Acknowledgments} 
This study was supported by the JSPS KAKENHI Grant Numbers JP18K03539, JP20H01854, JP21K03467, JP21K13879,  JP23H04870. The synchrotron radiation experiments at KEK were performed with the approval of the
Photon Factory Program Advisory Committee (Proposal No. 2022G520).

\bibliography{apssamp.bib}

\begin{thebibliography}{32}
\expandafter\ifx\csname natexlab\endcsname\relax\def\natexlab#1{#1}\fi
\expandafter\ifx\csname bibnamefont\endcsname\relax
  \def\bibnamefont#1{#1}\fi
\expandafter\ifx\csname bibfnamefont\endcsname\relax
  \def\bibfnamefont#1{#1}\fi
\expandafter\ifx\csname citenamefont\endcsname\relax
  \def\citenamefont#1{#1}\fi
\expandafter\ifx\csname url\endcsname\relax
  \def\url#1{\texttt{#1}}\fi
\expandafter\ifx\csname urlprefix\endcsname\relax\def\urlprefix{URL }\fi
\providecommand{\bibinfo}[2]{#2}
\providecommand{\eprint}[2][]{\url{#2}}

\bibitem[{\citenamefont{M{\"u}hlbauer et~al.}(2009)\citenamefont{M{\"u}hlbauer,
  Binz, Jonietz, Pfleiderer, Rosch, Neubauer, Georgii, and
  B{\"o}ni}}]{muhlbauer2009skyrmion}
\bibinfo{author}{\bibfnamefont{S.}~\bibnamefont{M{\"u}hlbauer}},
  \bibinfo{author}{\bibfnamefont{B.}~\bibnamefont{Binz}},
  \bibinfo{author}{\bibfnamefont{F.}~\bibnamefont{Jonietz}},
  \bibinfo{author}{\bibfnamefont{C.}~\bibnamefont{Pfleiderer}},
  \bibinfo{author}{\bibfnamefont{A.}~\bibnamefont{Rosch}},
  \bibinfo{author}{\bibfnamefont{A.}~\bibnamefont{Neubauer}},
  \bibinfo{author}{\bibfnamefont{R.}~\bibnamefont{Georgii}}, \bibnamefont{and}
  \bibinfo{author}{\bibfnamefont{P.}~\bibnamefont{B{\"o}ni}},
  \bibinfo{journal}{Science} \textbf{\bibinfo{volume}{323}},
  \bibinfo{pages}{915} (\bibinfo{year}{2009}).

\bibitem[{\citenamefont{Togawa et~al.}(2012)\citenamefont{Togawa, Koyama,
  Takayanagi, Mori, Kousaka, Akimitsu, Nishihara, Inoue, Ovchinnikov, and
  Kishine}}]{togawa2012chiral}
\bibinfo{author}{\bibfnamefont{Y.}~\bibnamefont{Togawa}},
  \bibinfo{author}{\bibfnamefont{T.}~\bibnamefont{Koyama}},
  \bibinfo{author}{\bibfnamefont{K.}~\bibnamefont{Takayanagi}},
  \bibinfo{author}{\bibfnamefont{S.}~\bibnamefont{Mori}},
  \bibinfo{author}{\bibfnamefont{Y.}~\bibnamefont{Kousaka}},
  \bibinfo{author}{\bibfnamefont{J.}~\bibnamefont{Akimitsu}},
  \bibinfo{author}{\bibfnamefont{S.}~\bibnamefont{Nishihara}},
  \bibinfo{author}{\bibfnamefont{K.}~\bibnamefont{Inoue}},
  \bibinfo{author}{\bibfnamefont{A.}~\bibnamefont{Ovchinnikov}},
  \bibnamefont{and} \bibinfo{author}{\bibfnamefont{J.-i.}
  \bibnamefont{Kishine}}, \bibinfo{journal}{Phys. Rev. Lett.}
  \textbf{\bibinfo{volume}{108}}, \bibinfo{pages}{107202}
  (\bibinfo{year}{2012}).

\bibitem[{\citenamefont{Togawa et~al.}(2016)\citenamefont{Togawa, Kousaka,
  Inoue, and Kishine}}]{togawa2016symmetry}
\bibinfo{author}{\bibfnamefont{Y.}~\bibnamefont{Togawa}},
  \bibinfo{author}{\bibfnamefont{Y.}~\bibnamefont{Kousaka}},
  \bibinfo{author}{\bibfnamefont{K.}~\bibnamefont{Inoue}}, \bibnamefont{and}
  \bibinfo{author}{\bibfnamefont{J.-i.} \bibnamefont{Kishine}},
  \bibinfo{journal}{J. Phys. Soc. Jpn.} \textbf{\bibinfo{volume}{85}},
  \bibinfo{pages}{112001} (\bibinfo{year}{2016}).

\bibitem[{\citenamefont{Momma and Izumi}(2011)}]{momma2011vesta}
\bibinfo{author}{\bibfnamefont{K.}~\bibnamefont{Momma}} \bibnamefont{and}
  \bibinfo{author}{\bibfnamefont{F.}~\bibnamefont{Izumi}},
  \bibinfo{journal}{Journal of applied crystallography}
  \textbf{\bibinfo{volume}{44}}, \bibinfo{pages}{1272} (\bibinfo{year}{2011}).

\bibitem[{\citenamefont{Ohara et~al.}(2011)\citenamefont{Ohara, Yamashita,
  Mori, and Sakamoto}}]{ohara2011transport}
\bibinfo{author}{\bibfnamefont{S.}~\bibnamefont{Ohara}},
  \bibinfo{author}{\bibfnamefont{T.}~\bibnamefont{Yamashita}},
  \bibinfo{author}{\bibfnamefont{Y.}~\bibnamefont{Mori}}, \bibnamefont{and}
  \bibinfo{author}{\bibfnamefont{I.}~\bibnamefont{Sakamoto}},
  \bibinfo{journal}{J. Phys.: Conf. Ser.} \textbf{\bibinfo{volume}{273}},
  \bibinfo{pages}{012048} (\bibinfo{year}{2011}).

\bibitem[{\citenamefont{Matsumura et~al.}(2017)\citenamefont{Matsumura, Kita,
  Kubo, Yoshikawa, Michimura, Inami, Kousaka, Inoue, and
  Ohara}}]{matsumura2017chiral}
\bibinfo{author}{\bibfnamefont{T.}~\bibnamefont{Matsumura}},
  \bibinfo{author}{\bibfnamefont{Y.}~\bibnamefont{Kita}},
  \bibinfo{author}{\bibfnamefont{K.}~\bibnamefont{Kubo}},
  \bibinfo{author}{\bibfnamefont{Y.}~\bibnamefont{Yoshikawa}},
  \bibinfo{author}{\bibfnamefont{S.}~\bibnamefont{Michimura}},
  \bibinfo{author}{\bibfnamefont{T.}~\bibnamefont{Inami}},
  \bibinfo{author}{\bibfnamefont{Y.}~\bibnamefont{Kousaka}},
  \bibinfo{author}{\bibfnamefont{K.}~\bibnamefont{Inoue}}, \bibnamefont{and}
  \bibinfo{author}{\bibfnamefont{S.}~\bibnamefont{Ohara}}, \bibinfo{journal}{J.
  Phys. Soc. Jpn.} \textbf{\bibinfo{volume}{86}}, \bibinfo{pages}{124702}
  (\bibinfo{year}{2017}).

\bibitem[{\citenamefont{Ninomiya et~al.}(2017)\citenamefont{Ninomiya,
  Matsumoto, Nakamura, Kono, Kittaka, Sakakibara, Inoue, and
  Ohara}}]{ninomiya2017magnetic}
\bibinfo{author}{\bibfnamefont{H.}~\bibnamefont{Ninomiya}},
  \bibinfo{author}{\bibfnamefont{Y.}~\bibnamefont{Matsumoto}},
  \bibinfo{author}{\bibfnamefont{S.}~\bibnamefont{Nakamura}},
  \bibinfo{author}{\bibfnamefont{Y.}~\bibnamefont{Kono}},
  \bibinfo{author}{\bibfnamefont{S.}~\bibnamefont{Kittaka}},
  \bibinfo{author}{\bibfnamefont{T.}~\bibnamefont{Sakakibara}},
  \bibinfo{author}{\bibfnamefont{K.}~\bibnamefont{Inoue}}, \bibnamefont{and}
  \bibinfo{author}{\bibfnamefont{S.}~\bibnamefont{Ohara}}, \bibinfo{journal}{J.
  Phys. Soc. Jpn.} \textbf{\bibinfo{volume}{86}}, \bibinfo{pages}{124704}
  (\bibinfo{year}{2017}).

\bibitem[{\citenamefont{Ishii et~al.}(2018)\citenamefont{Ishii, Takezawa,
  Mizuno, Kamikawa, Ninomiya, Matsumoto, Ohara, Mitsumoto, and
  Suzuki}}]{ishii2018ferroquadrupolar}
\bibinfo{author}{\bibfnamefont{I.}~\bibnamefont{Ishii}},
  \bibinfo{author}{\bibfnamefont{K.}~\bibnamefont{Takezawa}},
  \bibinfo{author}{\bibfnamefont{T.}~\bibnamefont{Mizuno}},
  \bibinfo{author}{\bibfnamefont{S.}~\bibnamefont{Kamikawa}},
  \bibinfo{author}{\bibfnamefont{H.}~\bibnamefont{Ninomiya}},
  \bibinfo{author}{\bibfnamefont{Y.}~\bibnamefont{Matsumoto}},
  \bibinfo{author}{\bibfnamefont{S.}~\bibnamefont{Ohara}},
  \bibinfo{author}{\bibfnamefont{K.}~\bibnamefont{Mitsumoto}},
  \bibnamefont{and} \bibinfo{author}{\bibfnamefont{T.}~\bibnamefont{Suzuki}},
  \bibinfo{journal}{J. Phys. Soc. Jpn.} \textbf{\bibinfo{volume}{87}},
  \bibinfo{pages}{013602} (\bibinfo{year}{2018}).

\bibitem[{\citenamefont{Ishii et~al.}(2019)\citenamefont{Ishii, Takezawa,
  Mizuno, Kumano, Suzuki, Ninomiya, Mitsumoto, Umeo, Nakamura, and
  Ohara}}]{ishii2019anisotropic}
\bibinfo{author}{\bibfnamefont{I.}~\bibnamefont{Ishii}},
  \bibinfo{author}{\bibfnamefont{K.}~\bibnamefont{Takezawa}},
  \bibinfo{author}{\bibfnamefont{T.}~\bibnamefont{Mizuno}},
  \bibinfo{author}{\bibfnamefont{S.}~\bibnamefont{Kumano}},
  \bibinfo{author}{\bibfnamefont{T.}~\bibnamefont{Suzuki}},
  \bibinfo{author}{\bibfnamefont{H.}~\bibnamefont{Ninomiya}},
  \bibinfo{author}{\bibfnamefont{K.}~\bibnamefont{Mitsumoto}},
  \bibinfo{author}{\bibfnamefont{K.}~\bibnamefont{Umeo}},
  \bibinfo{author}{\bibfnamefont{S.}~\bibnamefont{Nakamura}}, \bibnamefont{and}
  \bibinfo{author}{\bibfnamefont{S.}~\bibnamefont{Ohara}},
  \bibinfo{journal}{Phys. Rev. B} \textbf{\bibinfo{volume}{99}},
  \bibinfo{pages}{075156} (\bibinfo{year}{2019}).

\bibitem[{\citenamefont{Tsukagoshi et~al.}(2022)\citenamefont{Tsukagoshi,
  Matsumura, Michimura, Inami, and Ohara}}]{tsukagoshi2022competition}
\bibinfo{author}{\bibfnamefont{M.}~\bibnamefont{Tsukagoshi}},
  \bibinfo{author}{\bibfnamefont{T.}~\bibnamefont{Matsumura}},
  \bibinfo{author}{\bibfnamefont{S.}~\bibnamefont{Michimura}},
  \bibinfo{author}{\bibfnamefont{T.}~\bibnamefont{Inami}}, \bibnamefont{and}
  \bibinfo{author}{\bibfnamefont{S.}~\bibnamefont{Ohara}},
  \bibinfo{journal}{Phys. Rev. B} \textbf{\bibinfo{volume}{105}},
  \bibinfo{pages}{014428} (\bibinfo{year}{2022}).

\bibitem[{\citenamefont{Sato et~al.}(2022)\citenamefont{Sato, Manako, Homma,
  Li, Okazaki, and Aoki}}]{sato2022new}
\bibinfo{author}{\bibfnamefont{Y.~J.} \bibnamefont{Sato}},
  \bibinfo{author}{\bibfnamefont{H.}~\bibnamefont{Manako}},
  \bibinfo{author}{\bibfnamefont{Y.}~\bibnamefont{Homma}},
  \bibinfo{author}{\bibfnamefont{D.}~\bibnamefont{Li}},
  \bibinfo{author}{\bibfnamefont{R.}~\bibnamefont{Okazaki}}, \bibnamefont{and}
  \bibinfo{author}{\bibfnamefont{D.}~\bibnamefont{Aoki}},
  \bibinfo{journal}{Physical Review Materials} \textbf{\bibinfo{volume}{6}},
  \bibinfo{pages}{104412} (\bibinfo{year}{2022}).

\bibitem[{\citenamefont{Kakihana et~al.}(2018)\citenamefont{Kakihana, Aoki,
  Nakamura, Honda, Nakashima, Amako, Nakamura, Sakakibara, Hedo, Nakama
  et~al.}}]{kakihana2018giant}
\bibinfo{author}{\bibfnamefont{M.}~\bibnamefont{Kakihana}},
  \bibinfo{author}{\bibfnamefont{D.}~\bibnamefont{Aoki}},
  \bibinfo{author}{\bibfnamefont{A.}~\bibnamefont{Nakamura}},
  \bibinfo{author}{\bibfnamefont{F.}~\bibnamefont{Honda}},
  \bibinfo{author}{\bibfnamefont{M.}~\bibnamefont{Nakashima}},
  \bibinfo{author}{\bibfnamefont{Y.}~\bibnamefont{Amako}},
  \bibinfo{author}{\bibfnamefont{S.}~\bibnamefont{Nakamura}},
  \bibinfo{author}{\bibfnamefont{T.}~\bibnamefont{Sakakibara}},
  \bibinfo{author}{\bibfnamefont{M.}~\bibnamefont{Hedo}},
  \bibinfo{author}{\bibfnamefont{T.}~\bibnamefont{Nakama}},
  \bibnamefont{et~al.}, \bibinfo{journal}{J. Phys. Soc. Jpn.}
  \textbf{\bibinfo{volume}{87}}, \bibinfo{pages}{023701}
  (\bibinfo{year}{2018}).

\bibitem[{\citenamefont{Kakihana et~al.}(2019)\citenamefont{Kakihana, Aoki,
  Nakamura, Honda, Nakashima, Amako, Takeuchi, Harima, Hedo, Nakama
  et~al.}}]{kakihana2019unique}
\bibinfo{author}{\bibfnamefont{M.}~\bibnamefont{Kakihana}},
  \bibinfo{author}{\bibfnamefont{D.}~\bibnamefont{Aoki}},
  \bibinfo{author}{\bibfnamefont{A.}~\bibnamefont{Nakamura}},
  \bibinfo{author}{\bibfnamefont{F.}~\bibnamefont{Honda}},
  \bibinfo{author}{\bibfnamefont{M.}~\bibnamefont{Nakashima}},
  \bibinfo{author}{\bibfnamefont{Y.}~\bibnamefont{Amako}},
  \bibinfo{author}{\bibfnamefont{T.}~\bibnamefont{Takeuchi}},
  \bibinfo{author}{\bibfnamefont{H.}~\bibnamefont{Harima}},
  \bibinfo{author}{\bibfnamefont{M.}~\bibnamefont{Hedo}},
  \bibinfo{author}{\bibfnamefont{T.}~\bibnamefont{Nakama}},
  \bibnamefont{et~al.}, \bibinfo{journal}{J. Phys. Soc. Jpn.}
  \textbf{\bibinfo{volume}{88}}, \bibinfo{pages}{094705}
  (\bibinfo{year}{2019}).

\bibitem[{\citenamefont{Tabata et~al.}(2019)\citenamefont{Tabata, Matsumura,
  Nakao, Michimura, Kakihana, Inami, Kaneko, Hedo, Nakama, and
  {\=O}nuki}}]{tabata2019magnetic}
\bibinfo{author}{\bibfnamefont{C.}~\bibnamefont{Tabata}},
  \bibinfo{author}{\bibfnamefont{T.}~\bibnamefont{Matsumura}},
  \bibinfo{author}{\bibfnamefont{H.}~\bibnamefont{Nakao}},
  \bibinfo{author}{\bibfnamefont{S.}~\bibnamefont{Michimura}},
  \bibinfo{author}{\bibfnamefont{M.}~\bibnamefont{Kakihana}},
  \bibinfo{author}{\bibfnamefont{T.}~\bibnamefont{Inami}},
  \bibinfo{author}{\bibfnamefont{K.}~\bibnamefont{Kaneko}},
  \bibinfo{author}{\bibfnamefont{M.}~\bibnamefont{Hedo}},
  \bibinfo{author}{\bibfnamefont{T.}~\bibnamefont{Nakama}}, \bibnamefont{and}
  \bibinfo{author}{\bibfnamefont{Y.}~\bibnamefont{{\=O}nuki}},
  \bibinfo{journal}{J. Phys. Soc. Jpn.} \textbf{\bibinfo{volume}{88}},
  \bibinfo{pages}{093704} (\bibinfo{year}{2019}).

\bibitem[{\citenamefont{Sakakibara et~al.}(2019)\citenamefont{Sakakibara,
  Nakamura, Kittaka, Kakihana, Hedo, Nakama, and
  {\=O}nuki}}]{sakakibara2019fluctuation}
\bibinfo{author}{\bibfnamefont{T.}~\bibnamefont{Sakakibara}},
  \bibinfo{author}{\bibfnamefont{S.}~\bibnamefont{Nakamura}},
  \bibinfo{author}{\bibfnamefont{S.}~\bibnamefont{Kittaka}},
  \bibinfo{author}{\bibfnamefont{M.}~\bibnamefont{Kakihana}},
  \bibinfo{author}{\bibfnamefont{M.}~\bibnamefont{Hedo}},
  \bibinfo{author}{\bibfnamefont{T.}~\bibnamefont{Nakama}}, \bibnamefont{and}
  \bibinfo{author}{\bibfnamefont{Y.}~\bibnamefont{{\=O}nuki}},
  \bibinfo{journal}{J. Phys. Soc. Jpn.} \textbf{\bibinfo{volume}{88}},
  \bibinfo{pages}{093701} (\bibinfo{year}{2019}).

\bibitem[{\citenamefont{Topertser et~al.}(2019)\citenamefont{Topertser,
  Martyniak, Muts, Tokaychuk, and Gladyshevskii}}]{topertser2019crystal}
\bibinfo{author}{\bibfnamefont{V.}~\bibnamefont{Topertser}},
  \bibinfo{author}{\bibfnamefont{R.-I.} \bibnamefont{Martyniak}},
  \bibinfo{author}{\bibfnamefont{N.}~\bibnamefont{Muts}},
  \bibinfo{author}{\bibfnamefont{Y.}~\bibnamefont{Tokaychuk}},
  \bibnamefont{and}
  \bibinfo{author}{\bibfnamefont{R.}~\bibnamefont{Gladyshevskii}},
  \bibinfo{journal}{Chem. Met. Alloys} \textbf{\bibinfo{volume}{12}},
  \bibinfo{pages}{21} (\bibinfo{year}{2019}).

\bibitem[{\citenamefont{Nakamura et~al.}(2020)\citenamefont{Nakamura, Inukai,
  Asaka, Yamaura, and Ohara}}]{nakamura2020enantiopure}
\bibinfo{author}{\bibfnamefont{S.}~\bibnamefont{Nakamura}},
  \bibinfo{author}{\bibfnamefont{J.}~\bibnamefont{Inukai}},
  \bibinfo{author}{\bibfnamefont{T.}~\bibnamefont{Asaka}},
  \bibinfo{author}{\bibfnamefont{J.-i.} \bibnamefont{Yamaura}},
  \bibnamefont{and} \bibinfo{author}{\bibfnamefont{S.}~\bibnamefont{Ohara}},
  \bibinfo{journal}{J. Phys. Soc. Jpn.} \textbf{\bibinfo{volume}{89}},
  \bibinfo{pages}{104005} (\bibinfo{year}{2020}).

\bibitem[{\citenamefont{Ishida et~al.}(1985)\citenamefont{Ishida, Endoh,
  Mitsuda, Ishikawa, and Tanaka}}]{ishida1985crystal}
\bibinfo{author}{\bibfnamefont{M.}~\bibnamefont{Ishida}},
  \bibinfo{author}{\bibfnamefont{Y.}~\bibnamefont{Endoh}},
  \bibinfo{author}{\bibfnamefont{S.}~\bibnamefont{Mitsuda}},
  \bibinfo{author}{\bibfnamefont{Y.}~\bibnamefont{Ishikawa}}, \bibnamefont{and}
  \bibinfo{author}{\bibfnamefont{M.}~\bibnamefont{Tanaka}},
  \bibinfo{journal}{J. Phys. Soc. Jpn.} \textbf{\bibinfo{volume}{54}},
  \bibinfo{pages}{2975} (\bibinfo{year}{1985}).

\bibitem[{\citenamefont{Yamasaki et~al.}(2007)\citenamefont{Yamasaki, Sagayama,
  Goto, Matsuura, Hirota, Arima, and Tokura}}]{yamasaki2007electric}
\bibinfo{author}{\bibfnamefont{Y.}~\bibnamefont{Yamasaki}},
  \bibinfo{author}{\bibfnamefont{H.}~\bibnamefont{Sagayama}},
  \bibinfo{author}{\bibfnamefont{T.}~\bibnamefont{Goto}},
  \bibinfo{author}{\bibfnamefont{M.}~\bibnamefont{Matsuura}},
  \bibinfo{author}{\bibfnamefont{K.}~\bibnamefont{Hirota}},
  \bibinfo{author}{\bibfnamefont{T.}~\bibnamefont{Arima}}, \bibnamefont{and}
  \bibinfo{author}{\bibfnamefont{Y.}~\bibnamefont{Tokura}},
  \bibinfo{journal}{Phys. Rev. Lett.} \textbf{\bibinfo{volume}{98}},
  \bibinfo{pages}{147204} (\bibinfo{year}{2007}).

\bibitem[{\citenamefont{Sutter et~al.}(1997)\citenamefont{Sutter, Gr{\"u}bel,
  Vettier, De~Bergevin, Stunault, Gibbs, and Giles}}]{sutter1997helicity}
\bibinfo{author}{\bibfnamefont{C.}~\bibnamefont{Sutter}},
  \bibinfo{author}{\bibfnamefont{G.}~\bibnamefont{Gr{\"u}bel}},
  \bibinfo{author}{\bibfnamefont{C.}~\bibnamefont{Vettier}},
  \bibinfo{author}{\bibfnamefont{F.}~\bibnamefont{De~Bergevin}},
  \bibinfo{author}{\bibfnamefont{A.}~\bibnamefont{Stunault}},
  \bibinfo{author}{\bibfnamefont{D.}~\bibnamefont{Gibbs}}, \bibnamefont{and}
  \bibinfo{author}{\bibfnamefont{C.}~\bibnamefont{Giles}},
  \bibinfo{journal}{Phys. Rev. B} \textbf{\bibinfo{volume}{55}},
  \bibinfo{pages}{954} (\bibinfo{year}{1997}).

\bibitem[{\citenamefont{Fabrizi et~al.}(2009)\citenamefont{Fabrizi, Walker,
  Paolasini, de~Bergevin, Boothroyd, Prabhakaran, and
  McMorrow}}]{PhysRevLett.102.237205}
\bibinfo{author}{\bibfnamefont{F.}~\bibnamefont{Fabrizi}},
  \bibinfo{author}{\bibfnamefont{H.~C.} \bibnamefont{Walker}},
  \bibinfo{author}{\bibfnamefont{L.}~\bibnamefont{Paolasini}},
  \bibinfo{author}{\bibfnamefont{F.}~\bibnamefont{de~Bergevin}},
  \bibinfo{author}{\bibfnamefont{A.~T.} \bibnamefont{Boothroyd}},
  \bibinfo{author}{\bibfnamefont{D.}~\bibnamefont{Prabhakaran}},
  \bibnamefont{and} \bibinfo{author}{\bibfnamefont{D.~F.}
  \bibnamefont{McMorrow}}, \bibinfo{journal}{Phys. Rev. Lett.}
  \textbf{\bibinfo{volume}{102}}, \bibinfo{pages}{237205}
  (\bibinfo{year}{2009}).

\bibitem[{\citenamefont{Sagayama et~al.}(2010)\citenamefont{Sagayama, Abe,
  Taniguchi, Arima, Yamasaki, Okuyama, Tokura, Sakai, Morita, Komesu
  et~al.}}]{sagayama2010observation}
\bibinfo{author}{\bibfnamefont{H.}~\bibnamefont{Sagayama}},
  \bibinfo{author}{\bibfnamefont{N.}~\bibnamefont{Abe}},
  \bibinfo{author}{\bibfnamefont{K.}~\bibnamefont{Taniguchi}},
  \bibinfo{author}{\bibfnamefont{T.-h.} \bibnamefont{Arima}},
  \bibinfo{author}{\bibfnamefont{Y.}~\bibnamefont{Yamasaki}},
  \bibinfo{author}{\bibfnamefont{D.}~\bibnamefont{Okuyama}},
  \bibinfo{author}{\bibfnamefont{Y.}~\bibnamefont{Tokura}},
  \bibinfo{author}{\bibfnamefont{S.}~\bibnamefont{Sakai}},
  \bibinfo{author}{\bibfnamefont{T.}~\bibnamefont{Morita}},
  \bibinfo{author}{\bibfnamefont{T.}~\bibnamefont{Komesu}},
  \bibnamefont{et~al.}, \bibinfo{journal}{J. Phys. Soc. Jpn.}
  \textbf{\bibinfo{volume}{79}}, \bibinfo{pages}{043711}
  (\bibinfo{year}{2010}).

\bibitem[{\citenamefont{Kishine et~al.}(2005)\citenamefont{Kishine, Inoue, and
  Yoshida}}]{kishine2005synthesis}
\bibinfo{author}{\bibfnamefont{J.-i.} \bibnamefont{Kishine}},
  \bibinfo{author}{\bibfnamefont{K.}~\bibnamefont{Inoue}}, \bibnamefont{and}
  \bibinfo{author}{\bibfnamefont{Y.}~\bibnamefont{Yoshida}},
  \bibinfo{journal}{Progress of Theoretical Physics Supplement}
  \textbf{\bibinfo{volume}{159}}, \bibinfo{pages}{82} (\bibinfo{year}{2005}).

\bibitem[{\citenamefont{Kousaka et~al.}(2007)\citenamefont{Kousaka, Yano,
  Kishine, Yoshida, Inoue, Kikuchi, and Akimitsu}}]{kousaka2007chiral}
\bibinfo{author}{\bibfnamefont{Y.}~\bibnamefont{Kousaka}},
  \bibinfo{author}{\bibfnamefont{S.-i.} \bibnamefont{Yano}},
  \bibinfo{author}{\bibfnamefont{J.-i.} \bibnamefont{Kishine}},
  \bibinfo{author}{\bibfnamefont{Y.}~\bibnamefont{Yoshida}},
  \bibinfo{author}{\bibfnamefont{K.}~\bibnamefont{Inoue}},
  \bibinfo{author}{\bibfnamefont{K.}~\bibnamefont{Kikuchi}}, \bibnamefont{and}
  \bibinfo{author}{\bibfnamefont{J.}~\bibnamefont{Akimitsu}},
  \bibinfo{journal}{J. Phys. Soc. Jpn.} \textbf{\bibinfo{volume}{76}},
  \bibinfo{pages}{123709} (\bibinfo{year}{2007}).

\bibitem[{\citenamefont{Hidaka et~al.}(2020)\citenamefont{Hidaka, Mizuuchi,
  Hayasaka, Yanagisawa, Ohara, and Amitsuka}}]{hidaka2020helical}
\bibinfo{author}{\bibfnamefont{H.}~\bibnamefont{Hidaka}},
  \bibinfo{author}{\bibfnamefont{K.}~\bibnamefont{Mizuuchi}},
  \bibinfo{author}{\bibfnamefont{E.}~\bibnamefont{Hayasaka}},
  \bibinfo{author}{\bibfnamefont{T.}~\bibnamefont{Yanagisawa}},
  \bibinfo{author}{\bibfnamefont{J.}~\bibnamefont{Ohara}}, \bibnamefont{and}
  \bibinfo{author}{\bibfnamefont{H.}~\bibnamefont{Amitsuka}},
  \bibinfo{journal}{Phys. Rev. B} \textbf{\bibinfo{volume}{102}},
  \bibinfo{pages}{174408} (\bibinfo{year}{2020}).

\bibitem[{\citenamefont{Bouvier et~al.}(1991)\citenamefont{Bouvier,
  Lethuillier, and Schmitt}}]{PhysRevB.43.13137}
\bibinfo{author}{\bibfnamefont{M.}~\bibnamefont{Bouvier}},
  \bibinfo{author}{\bibfnamefont{P.}~\bibnamefont{Lethuillier}},
  \bibnamefont{and} \bibinfo{author}{\bibfnamefont{D.}~\bibnamefont{Schmitt}},
  \bibinfo{journal}{Phys. Rev. B} \textbf{\bibinfo{volume}{43}},
  \bibinfo{pages}{13137} (\bibinfo{year}{1991}).

\bibitem[{\citenamefont{Blanco et~al.}(1991)\citenamefont{Blanco, Gignoux, and
  Schmitt}}]{PhysRevB.43.13145}
\bibinfo{author}{\bibfnamefont{J.~A.} \bibnamefont{Blanco}},
  \bibinfo{author}{\bibfnamefont{D.}~\bibnamefont{Gignoux}}, \bibnamefont{and}
  \bibinfo{author}{\bibfnamefont{D.}~\bibnamefont{Schmitt}},
  \bibinfo{journal}{Phys. Rev. B} \textbf{\bibinfo{volume}{43}},
  \bibinfo{pages}{13145} (\bibinfo{year}{1991}).

\bibitem[{\citenamefont{Hannon et~al.}(1988)\citenamefont{Hannon, Trammell,
  Blume, and Gibbs}}]{PhysRevLett.61.1245}
\bibinfo{author}{\bibfnamefont{J.~P.} \bibnamefont{Hannon}},
  \bibinfo{author}{\bibfnamefont{G.~T.} \bibnamefont{Trammell}},
  \bibinfo{author}{\bibfnamefont{M.}~\bibnamefont{Blume}}, \bibnamefont{and}
  \bibinfo{author}{\bibfnamefont{D.}~\bibnamefont{Gibbs}},
  \bibinfo{journal}{Phys. Rev. Lett.} \textbf{\bibinfo{volume}{61}},
  \bibinfo{pages}{1245} (\bibinfo{year}{1988}).

\bibitem[{\citenamefont{Lovesey et~al.}(2005)\citenamefont{Lovesey, Balcar,
  Knight, and Rodr{\'\i}guez}}]{lovesey2005electronic}
\bibinfo{author}{\bibfnamefont{S.~W.} \bibnamefont{Lovesey}},
  \bibinfo{author}{\bibfnamefont{E.}~\bibnamefont{Balcar}},
  \bibinfo{author}{\bibfnamefont{K.}~\bibnamefont{Knight}}, \bibnamefont{and}
  \bibinfo{author}{\bibfnamefont{J.~F.} \bibnamefont{Rodr{\'\i}guez}},
  \bibinfo{journal}{Physics Reports} \textbf{\bibinfo{volume}{411}},
  \bibinfo{pages}{233} (\bibinfo{year}{2005}).

\bibitem[{Lov()}]{Lovesey}
\bibinfo{note}{S. W. Lovesey and S. P. Collins, \it{X-ray Scattering and
  Absorption by Magnetic Materials} \rm{(Oxford University Press, New York,
  1996).}}

\bibitem[{\citenamefont{Zheludev et~al.}(1997)\citenamefont{Zheludev, Maslov,
  Shirane, Sasago, Koide, and Uchinokura}}]{zheludev1997field}
\bibinfo{author}{\bibfnamefont{A.}~\bibnamefont{Zheludev}},
  \bibinfo{author}{\bibfnamefont{S.}~\bibnamefont{Maslov}},
  \bibinfo{author}{\bibfnamefont{G.}~\bibnamefont{Shirane}},
  \bibinfo{author}{\bibfnamefont{Y.}~\bibnamefont{Sasago}},
  \bibinfo{author}{\bibfnamefont{N.}~\bibnamefont{Koide}}, \bibnamefont{and}
  \bibinfo{author}{\bibfnamefont{K.}~\bibnamefont{Uchinokura}},
  \bibinfo{journal}{Physical review letters} \textbf{\bibinfo{volume}{78}},
  \bibinfo{pages}{4857} (\bibinfo{year}{1997}).

\bibitem[{\citenamefont{Ninomiya et~al.}(2018)\citenamefont{Ninomiya, Sato,
  Matsumoto, Moyoshi, Nakao, Ohishi, Kousaka, Akimitsu, Inoue, and
  Ohara}}]{ninomiya2018neutron}
\bibinfo{author}{\bibfnamefont{H.}~\bibnamefont{Ninomiya}},
  \bibinfo{author}{\bibfnamefont{T.}~\bibnamefont{Sato}},
  \bibinfo{author}{\bibfnamefont{Y.}~\bibnamefont{Matsumoto}},
  \bibinfo{author}{\bibfnamefont{T.}~\bibnamefont{Moyoshi}},
  \bibinfo{author}{\bibfnamefont{A.}~\bibnamefont{Nakao}},
  \bibinfo{author}{\bibfnamefont{K.}~\bibnamefont{Ohishi}},
  \bibinfo{author}{\bibfnamefont{Y.}~\bibnamefont{Kousaka}},
  \bibinfo{author}{\bibfnamefont{J.}~\bibnamefont{Akimitsu}},
  \bibinfo{author}{\bibfnamefont{K.}~\bibnamefont{Inoue}}, \bibnamefont{and}
  \bibinfo{author}{\bibfnamefont{S.}~\bibnamefont{Ohara}},
  \bibinfo{journal}{Physica B: Condensed Matter}
  \textbf{\bibinfo{volume}{536}}, \bibinfo{pages}{392} (\bibinfo{year}{2018}).

\end{thebibliography}

\end{document}